# Flow-Velocity-Dependent Transition of Anisotropic Crack Patterns in CaCO$_3$ Paste

Yuri Akiba and Hiroyuki Shima*

*Department of Environmental Sciences, University of Yamanashi, 4-4-37 Takeda, Kofu 400-8510, Japan*

We investigate the desiccation crack patterns on the surface of a drying paste made of calcium carbonate (CaCO$_3$) powder and distilled water. Forced vibration of the CaCO$_3$ paste prior to drying results in an anisotropic crack pattern, in which many long cracks develop along a specific preferred direction. We reveal that the preferred direction changes from perpendicular to parallel to the vibration direction at the threshold velocity of vibration. The transition is attributed to the reorientation of constituent particles subjected to the forced oscillatory flow of fluid in the paste.

## 1.  Introduction

Desiccation cracks are widely observed on surfaces of granular pastes. A paste is a mixture of liquid and insoluble grains; upon drying, the liquid content evaporates from the air-exposed surface, causing volume shrinkage of the grain-liquid mixture followed by surface cracking.[1,2] The fractured surface often exhibits an isotropic polygonal pattern similar to those that develop on the dried surfaces of mud[2–4] and paint.[5,6] The patterns are characterized by cracks with different lengths and directions, splitting the entire surface of the fractured medium into many polygonal cells.[7–11]

Although desiccation cracks have been widely studied, it is still difficult to predict the direction or position of crack advancement on a drying surface from actual experimental results, even from a statistical viewpoint. The difficulty stems partly from the fact that the crack behavior is governed by complex interactions among microscopic particles.[12,13] In fact, the internal structure of granular pastes is spatially inhomogeneous because of the wide dispersion both in the spatial distribution of constituent particles and in their particle size distribution; in addition, the spatial inhomogeneity in the drying rate of the paste also contributes to the complex interactions.[14] Furthermore, as a crack progresses through a paste, the local stress field in the vicinity of the crack changes with time,[15] which makes the problem more difficult. On the other hand, it has been reported that the desiccation crack pattern in some pastes can be controlled to a degree by applying an oscillatory force in the parallel direction to the thin paste before drying; this is the so-called memory effect of paste.[16] The memory effect occurs when a paste is vibrated horizontally for a few minutes before the drying process.[17,18] This would make it possible to control the direction and position of crack advancement.

A paste of calcium carbonate (CaCO$_3$) is a typical material exhibiting the memory effect.[19] Figure 1 shows experimental results of CaCO$_3$ pastes dried completely. It is known that CaCO$_3$ pastes exhibit two types of memory effects. One is the memory effect of "shaking", characterized by many long surface cracks propagating in a direction perpendicular to the vibration direction of a thin container in which the paste is poured; see Fig. 1(a). The other effect is called the memory effect of "flow". This effect is manifested as many long cracks that propagate parallel to the direction of flow induced by the horizontal vibration of the container; see Fig. 1(b). Here, the term flow refers to the vibration-induced solvent flow that occurs around the constituent particles of the paste. The two types of memory effects result in anisotropic crack patterns that are clearly different from the ordinary random, isotropic crack pattern shown in Fig. 1(c). The memory effects have academic significance in the apparently counterintuitive fact that the global geometry of desiccation crack patterns can be altered merely by imposing a horizontal vibration.

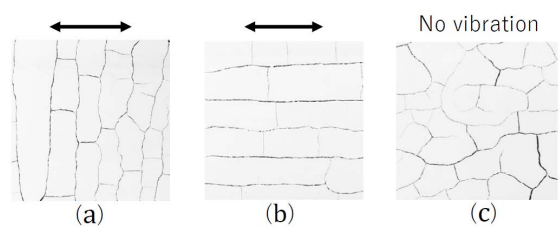

Fig. 1.  Experimental results of desiccation cracking in CaCO$_3$ paste: a) anisotropic crack pattern indicating memory effect of shaking, b) anisotropic crack pattern indicating memory effect of flow, and c) isotropic crack pattern consisting of randomly oriented cracks. The vertical arrays indicate the vibration direction of the container in which the paste was stored.

In an earlier study, the mechanism behind the memory effect of shaking was explained using elastoplastic theories.[20] The theory suggests that the memory effect of shaking results from the anisotropy in the residual stresses due to the plastic deformation of a paste. In addition, a nonlinear analysis, wherein a nonlinear effect is introduced into the elastoplastic model, has been proposed.[21] These theories are consistent with experimental observations, thus making them suitable to describe the memory effect of shaking. In contrast, the mechanism behind the memory effect of flow has not been fully explained. A plausible idea[19] is based on flow-induced anisotropy in the particle configuration. When a solvent flows through the gap between constituent particles of a paste, the particles form a network structure because of the hydrodynamic attractive interaction. As the flow continues, the overall network structure becomes elongated, thereby orienting many chains of the particles in the flow direction. In fact, a numerical simulation showed that the hydrodynamic interaction may lead to an anisotropic chain structure of the particles in col-loidal suspensions.[22] If the constituent particles in the paste form a chainlike network structure, the flow velocity around the particles driven by the horizontal vibration of the container is expected to be responsible for the degree of elongation in the network. Intuitively, a strong flow will enhance the elongation and thus promote the occurrence of the memory effect of flow. However, there has been no experimental observation that supports this plausible idea.

*Corresponding Author (H. Shima): hshima@yamanashi.ac.jp

The present work is aimed at verifying the conjecture that the flow velocity governs the mechanism behind the memory effect of flow in pastes. We conduct drying crack experiments on $CaCO_3$ paste to examine the correlation between the flow velocity and the degree of anisotropy in the crack pattern as a manifestation of the memory effect. The systematic tuning of the flow velocity shows that the memory effect is enhanced with increase in the flow velocity, thus supporting the aforementioned theoretical scenario.

## 2. Materials and Methods

### 2.1 Sample preparation

We used a paste of $CaCO_3$ as the experimental material to investigate the memory effect of flow. It is composed of white, micrometer-size particles that are insoluble in water. In our actual experiments, we mixed 95.65 g of $CaCO_3$ powder (Hayashi Pure Chemical Industries) and 104.35 g of distilled water in a container. The solid volume fraction of the paste was set to 25%, at which the paste has been reported to exhibit the memory effect of flow.[19] It is known that $CaCO_3$ powder is composed of electrically charged particles. Because the Coulombic repulsive interaction between the charged particles will hinder the network structure formation, we added sodium chloride (Nihonkaisui) to the $CaCO_3$ paste as an electrolyte that neutralizes the paste. The added amount of sodium chloride was set to 0.61 g (0.1 mol/L for the paste), which was sufficient to neutralize the calcium carbonate.

### 2.2 Measurement method

The paste was poured into a plastic, rectangular container. The longer side, narrow side, and depth of the container were 215, 153, and 39 mm, respectively. The container was fixed on a horizontal plate on a shaker (TOKYO RIKAKIKAI, MMS3010). Thereafter, the container was vibrated horizontally for 60 s. Figure 2 shows a schematic of the experimental setup.

The vibration direction was parallel to the longer side of the container. The flow velocity was tuned by controlling the amplitude and frequency of the vibration in a separate manner; the two quantities were varied as 10, 20, and 30 mm and 1.33, 2.17, and 3.00 Hz, respectively. In addition, we examined the case of a significantly lower flow velocity by setting a 10 mm in amplitude and 0.67 Hz frequency. After the vibration, the samples were left to rest and dried completely for 24 h in a room maintained at 28 °C. The images of the polygonal cracks after drying were analyzed using the image processing software ArcGIS,[23] which is a geographic information system tool employed to analyze geographic data.

The flow velocity was controlled by varying the amplitude ($r$) and frequency ($f$) of the horizontal vibration of the container. The vibration was in the form of a simple harmonic oscillation. Accordingly, the velocity of vibration $V$ is given

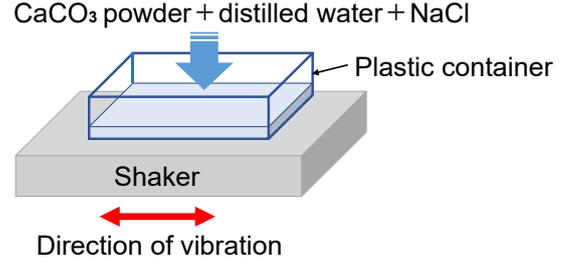

**Fig. 2.** (Color online) Diagram of experimental setup for paste sample vibration.

by
$$V = r\omega \cos \omega t, \qquad (1)$$

where $\omega (= 2\pi f)$ is the angular velocity and $t$ is the duration for which the oscillation is applied. Hence, the maximum velocity of oscillation ($V_{max}$) is represented by $V_{max} = r\omega$.

### 2.3 Order parameter

The geographic data of crack patterns was quantified using the orientation order parameter defined as follows. We assume that a completely dried surface of a colloidal suspension is divided into many polygonal cells by $N$ cracks with different crack lengths and orientations. In the following, $\ell_i$ represents the length of the $i$th crack ($1 \le i \le N$) and $\theta_i$ denotes the angle between the direction of the $i$th crack and the direction along which the sample is vibrated. Figure 3 shows a schematic defining $\ell_i$ and $\theta_i$. In particular, if a crack is oriented in the vibration direction, we have $\theta_i = 0$; otherwise, if a crack is perpendicular to the vibration direction, we have $\theta_i = \pi/2$. In general, $\theta_i$ takes an intermediate value ranging from 0 to $\pi/2$. Now, we define the order parameter $O$ of the system as

$$O \equiv 2C_2 - 1, \quad C_m = \frac{\sum_{i=1}^{N} \ell_i^m \cos^2 \theta_i}{\sum_{i=1}^{N} \ell_i^m}. \qquad (2)$$

The reason for our choice of $m = 2$ in Eq. (2) will be given soon. Alternatively, $O$, expressed in Eq. (2), can be rewritten by the sum of $\cos^2 \theta_i$ with weight $w_i^{(m)}$ as

$$O \equiv 2 \left( \sum_{i=1}^{N} w_i^{(m)} \cos^2 \theta_i \right) - 1, \quad w_i^{(m)} = \frac{\ell_i^m}{\sum_{i=1}^{N} \ell_i^m}. \qquad (3)$$

Below are the values of $O$ for three extreme cases.

**i)** When all the cracks are oriented in the vibration direction, we have $\theta_i \equiv 0$ for arbitrary $i$. Substituting this into Eq. (2), we obtain $C_2 = 1$ and $O = 1$. This case corresponds to the perfect realization of the memory effect of flow.

**ii)** When all the cracks are oriented in a direction perpendicular to the vibration direction, we have $\theta_i \equiv \pi/2$ for arbitrary $i$. Substituting this into Eq. (2), we obtain $C_2 = 0$ and $O = -1$. This case corresponds to the perfect realization of the memory effect of shaking.

**iii)** If all the cracks are randomly oriented but have an identical length $\ell_0$, $C_2$ is obtained as

$$C_2 = \frac{\ell_0^2 \sum_{i=1}^{N} \cos^2 \theta}{N \ell_0^2} = \frac{1}{N} \sum_{i=1}^{N} \cos^2 \theta. \quad (4)$$

For a sufficiently large $N$, the sum on the right of Eq. (4) is replaced by an integral; thus, $C_2$ is approximated as

$$C_2 \simeq \frac{\int_0^\pi \cos^2 \theta \, d\theta}{\int_0^\pi d\theta} = \frac{1}{\pi} \int_0^\pi \cos^2 \theta \, d\theta = \frac{1}{2}. \quad (5)$$

This leads to $O = 0$ for a randomly oriented case.

It follows from i)-iii) that when the order parameter takes a positive (negative) value, it indicates the occurrence of the memory effect of flow (shaking).

We should mention why we set $m = 2$ in the definitions of $C_m$ and $O$ in Eq. (2). It is important that the integer $m$ be greater than unity; otherwise, the defined $O$ is unable to characterize the degree of anisotropy when the total length of cracks with $\theta_i = 0$ is equal to that with $\theta_i = \pi/2$.

The need for $m > 1$ is explained by considering the brick-wall-type crack pattern depicted in Fig. 4. It consists of $n$ long cracks with length $\ell_l$ and $n(n + 1)$ short cracks with length $\ell_s$. The long cracks are equispaced, thus dividing the whole square sample with linear dimension $L$ into $n + 1$ stripes with length $L$ and width $L/(n + 1)$. Every stripe is further divided by $n$ short cracks. We thus have $\ell_l = L$ and $\ell_s = L/(n + 1)$, which imply that the total length of the long cracks is equal to that of short cracks (i.e., $nL$).

Now we evaluate $O$ and $C_m$ for the two kinds of brick-wall patterns shown in Figs. 4(a) and 4(b). If we set $m = 1$, both patterns give $C_1 = 1/2$ and $O = 0$, which means that the defined $O$ is unable to distinguish the two brick-wall patterns. The problem is overcome by setting $m > 1$. If we set $m = 2$, for instance, we have $C_2 = (n + 1)/(n + 2)$, and $O = n/(n + 2) > 0$ for Fig. 4(a) and $C_2 = 1/(n + 2)$ and $O = -n/(n + 2) < 0$ for Fig. 4(b). Note that the sign of $O$ given above indicates which class of memory effects (shaking or flow) occurs. In addition, the value of $O$ approaches $+1$ (or $-1$) with increasing number of long cracks that are parallel (perpendicular) to the vibration direction. This approaching behavior is consistent with our criteron that the directions of long cracks are dominant in the determination of the class of memory effects. The above discussion holds true for every integer $m$ greater than unity. Among the choices, we set $m = 2$ in the present work for simplicity.

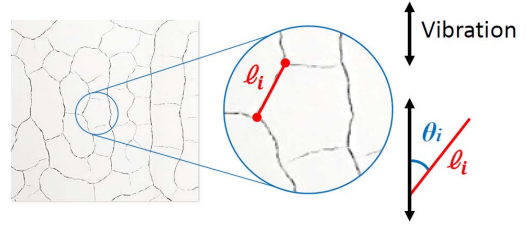

**Fig. 3.** (Color online) Schematic of the length of the $i$th crack ($\ell_i$) and the relative angle between the $i$th crack direction and the vibration direction of the sample ($\theta_i$).

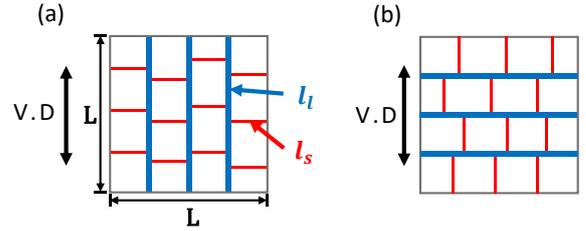

**Fig. 4.** (Color online) Schematics of a brick-wall-type crack pattern composed of $n$ long cracks with length $\ell_l$ and $n(n + 1)$ short cracks with length $\ell_s$. The case of $n = 3$ is depicted as an example. a) Pattern in which all long cracks are parallel to the vibration direction (V.D). b) Pattern in which all long cracks are perpendicular to the V.D.

## 3. Results

Our basic assumption is that the flow velocity governs the mechanism behind the memory effect of flow in the paste. There are two suggesting reasons. One is that solvent flow in the paste drives the reorientation of constituent particles; the higher the velocity of the solvent flow is, the easier it is for the particles to form a network structure. The other reason is that a high solvent flow velocity will promote the frequency of particle collisions. These motivated us to examine the correlation between the flow velocity and the orientational order in crack patterns.

Figure 5 demonstrates the dependence of the order parameter $O$ on the maximum velocity of vibration $V_{\max}(= r\omega)$. The number of samples is three for each data plot. The upper and lower ends of each error bar represent the maximum and minimum values of $O$, respectively, evaluated from three samples. It is clear from Fig. 5 that at sufficiently high $r\omega$ ($\simeq 0.5$ m/s), $O$ takes a positive value ($\simeq 0.4$). With decreasing $r\omega$ from 0.5 m/s, $O$ decreases and its sign changes from positive to negative at $r\omega=0.3$ m/s or slightly above. This sign inversion indicates that the value of 0.3 m/s is the transition point from the shaking phase to the flow phase. With a further decrease in $r\omega$, $O$ continues to decrease until it reaches a minimum at $r\omega \simeq 0.1$ m/s, followed by a sudden increase toward the origin. This convergence to the origin is intuitively understood because at a significantly small $r\omega$, the weak solvent flow provides insufficient force to trigger the reorientation of constituent particles. In fact, we visually confirmed in experiments that the crack patterns obtained near zero velocity (as well as those near the transition point) exhibited no anisotropy, as shown in Fig. 1(c).

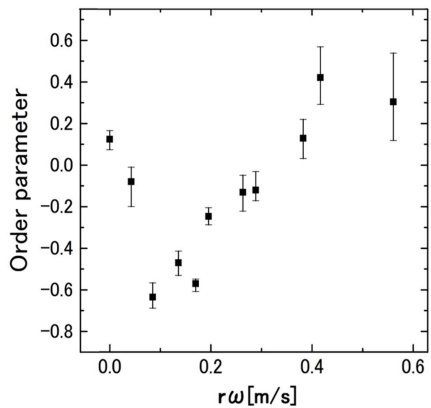

**Fig. 5.** Dependence of the order parameter on the flow velocity $r\omega$. The number of samples is three for each data plot. The upper (lower) end of the error bar represents the maximum (minimum) value of the order parameter evaluated from three samples.

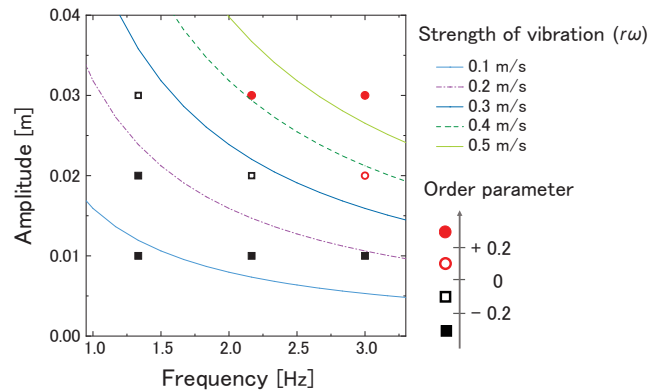

**Fig. 6.** (Color online) Phase diagram of the order parameter, amplitude, and frequency. The data points marked with red circles are positive values of the order parameter, indicating the occurrence of the memory effect of flow. The negative values are marked by black squares and indicate the occurrence of the memory effect of shaking. In either case, the magnitude of the order parameter is represented using different symbols (solid or open) depending on whether the absolute value is greater or lower than 0.2. The curves depicted in the diagram are the contours along which the maximum vibration velocity $V_{\max}(= r\omega)$ remains constant, as listed in the legend.

Figure 6 shows the phase diagram of the order parameter distribution in the amplitude-frequency plane. The data points marked with red circles are positive values of $O$, indicating the occurrence of the memory effect of flow at the given amplitude and frequency. The negative values are marked by black squares and indicate the occurrence of the memory effect of shaking. In either case, the magnitude of $O$ is represented using different symbols (solid or open) depending on whether the absolute value is greater or lower than 0.2. It follows from the diagram that the memory effect of flow tends to occur in the upper-right region, whereas the memory effect of shaking occurs in the lower-left region. The curves depicted in the diagram are the contours along which the maximum vibration velocity $V_{\max}(= r\omega)$ remains constant, as listed in the legend. The contour line of 0.3 m/s serves as the threshold, beyond which the memory effect changes from the shaking type (below the threshold) to the flow type (above the threshold). In addition, it is inferred from the contour map that the vibration amplitude and frequency mutually contribute to the determination of which type of memory effect, i.e., flow or shaking, occurs at the given parameter values. The product of the two parameters (i.e., the maximum flow velocity) uniquely determines the type of memory effect to be observed.

## 4. Discussion

Our experimental results clearly show a strong correlation between the flow velocity and the orientation order induced by the memory effect of flow. With increasing flow velocity, the drying crack pattern undergoes a transition from a shaking phase to a flow phase, each of which exhibits completely different patterns with respect to the preferred direction of crack formation. To elucidate the physical origin of the transition, we consider how the fluid flow penetrating through the gaps between the particles affects their configuration from a microscopic viewpoint.

In hydrodynamics, under conditions of sufficiently low fluid velocity and uniform flow around the particles, the fluid force acting on a particle can be represented by

$$F = 6\pi\mu RV. \quad (6)$$

Here, $\mu$ represents the viscosity coefficient of water, $R$ is the particle radius, and $V$ denotes the flow velocity. This equation indicates that the fluid force increases linearly with the flow velocity. Equation (6) is expected to be applicable to the colloid particles of the paste used in this study, because the solid volume fraction of the paste is 25% and thus the spaces within which particles can move freely are sufficiently large.

Assuming that a fluid force, expressed by Eq. (6), acts on a particle, it is plausible that a high fluid velocity enhances the anisotropy in the particle configuration, thus generating a chain like structure that orients in the vibration direction. Figures 7(a) and 7(b) show the nucleation of such a flow-induced chain. If the direction of the particle pair does not coincide with that of the flow [e.g., perpendicular to each other, as shown in Fig. 7(a)], the oscillatory flow of the fluid exerts a force on the pair such that it reorients in a direction parallel to the flow direction [Fig. 7(b)]. The same reorientation due to the oscillatory flow will occur even in small aggregates having more than two particles, resulting in elongated short chains of the constituent particles. Once the chains are bonded to each other via repeated displacement and collision by the flow, a large-scale anisotropic network structure of particles is formed. The gap between adjacent chains is mechanically fragile and is likely to fracture under drying.

In the discussion so far, we assumed that the oscillatory flow of the solvent in the vicinity of particles gives rise to a force that pushes them back and forth. This is plausible, particularly when the solid volume fraction is sufficiently low. If the fraction is sufficiently high, there may be an alternative force responsible for the particle displacement; this force is transmitted from the side and bottom walls of the vibrated container to the particles. In such a dense paste, the vibrated walls exert a force on particles alongside the walls, then this force transmits through the paste inward via the bonding between particles. In this case, the force acting on particles is expressed by, instead of Eq. (6), the product of the particle mass and the acceleration of the horizontally vibrated container; as a consequence, the acceleration, not the flow velocity, may be the determining factor of the type of memory effect.[19]

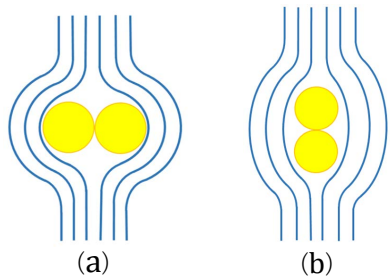

**Fig. 7.** (Color online) Orientation of particle pairs with respect to the flow direction: a) perpendicular and b) parallel.

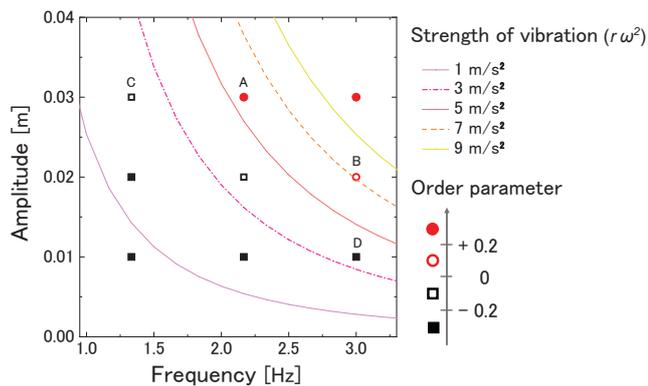

**Fig. 8.** (Color online) Duplication of Fig. 6 with the contour lines replaced with lines along which the horizontal acceleration of the container remains constant (as listed in the legend).

To test the alternative scenario mentioned above, we replot the phase diagram of Fig. 6 by replacing the contour lines with lines along which the horizontal acceleration of the container remains constant. The results are given in Fig. 8. It seems from the figure that the contour line of 5 m/s$^2$ may serve as the threshold above (below) which the order parameter tends to take a positive (negative) value. Compared with Fig. 6, nevertheless, there is a slight mismatch between the contour lines and the order parameter distribution in Fig. 8. For instance, points A and B in Fig. 8 are located to the left and right of the 7 m/s$^2$ line, respectively, which is inconsistent with the hypothesis of strong correlation between the acceleration and the orientational ordering. The same is true for points C and D. From this observation, it is fair to conclude that the flow velocity should be a determining factor of the type of memory effect rather than the acceleration under the present conditions. It is also inferred that in actual pastes, both the flow velocity and the acceleration contribute to the determination of which type of memory effect occurs (as well as the generation of the force acting on particles), while relative magnitudes of the contribution are expected to depend on the solid volume fraction in the pastes.

## 5. Conclusion

In this work, we experimentally obtained evidence of a strong correlation between the flow velocity and the orientation order with regard to anisotropic crack patterns in CaCO$_3$ paste. We presented a phase diagram with respect to the type of memory effect in the vibration amplitude-frequency plane. Furthermore, we demonstrated that the transition between the two types of memory effects occurs at a threshold flow velocity of approximately 0.3 m/s. The experimental findings support the plausible scenario that the memory effect of flow in pastes is due to the flow-induced formation of anisotropic chainlike structures.

Y. A. acknowledges the financial support from the Sasakawa Scientific Research Grant from Japan Science Society. The research was also supported by JSPS KAKENHI (Grant Nos. 16K00810 and 18H03818). Fruitful discussion with Professor Motohiro Sato is gratefully acknowledged.